	\newtheorem{problem}{Problem}
        \newcommand{\et}{\hspace{-0.08in}{\bf .}\hspace{0.1in}}
	\newcommand{\maps}{\colon}
	\renewcommand{\to}{\rightarrow}
        \newcommand{\vol}{{\rm vol}}
        \newcommand{\T}{{\cal T}}
        \newcommand{\tr}{{\rm tr}}
        
	\newcommand{\Fun}{{\rm Fun}}
	\newcommand{\Diff}{{\rm Diff}}
	\newcommand{\Aut}{{\rm Aut}}
        \newcommand{\ad}{{\rm ad}}
	
	\newcommand{\om}{\omega}
        \newcommand{\we}{\wedge}
	\newcommand{\tensor}{\otimes}
	\newcommand{\Con}{{\cal C}}
	\renewcommand{\H}{{\cal H}}
	\newcommand{\A}{{\cal A}}
	\newcommand{\G}{{\cal G}}
	\newcommand{\C}{{\bf C}}   
        \newcommand{\R}{{\bf R}}   
	\newcommand{\SO}{{\rm SO}}
	\newcommand{\SL}{{\rm SL}}
	\newcommand{\SU}{{\rm SU}}
	\newcommand{\U}{{\rm U}}

        \newcommand{\be}{\begin{equation}}
        \newcommand{\ee}{\end{equation}}
        \newcommand{\ba}{\begin{eqnarray}}
        \newcommand{\ea}{\end{eqnarray}}
	\newcommand{\ban}{\begin{eqnarray*}}
        \newcommand{\ean}{\end{eqnarray*}}
        \newcommand{\barr}{\begin{array}}
	\newcommand{\earr}{\end{array}}

	\documentstyle[12pt]{article}
\begin{document}
	\begin{center}
	{\bf Knots and Quantum Gravity:\\
          Progress and Prospects\\ }

	\vspace{0.5cm}
	{\em John C. Baez\\}
	\vspace{0.3cm}
	{\small Department of Mathematics \\
	University of California  \\
	Riverside CA 92521\\ }
	\vspace{0.3cm}
	{\small October 2, 1994\\ }
        \vspace{0.3 cm}
        {\small to appear in proceedings of the \\
        Seventh Marcel Grossman Meeting \\ }
	\vspace{0.5cm}
	\end{center}

\begin{abstract}

Recent work on the loop representation of quantum gravity has revealed
previously unsuspected connections between knot theory and quantum
gravity, or more generally, 3-dimensional topology and 4-dimensional
generally covariant physics.  We review how some of these relationships
arise from a `ladder of field theories' including quantum gravity and
$BF$ theory in 4 dimensions, Chern-Simons theory in 3 dimensions, and the
$G/G$ gauged WZW model in 2 dimensions.  We also describe the relation between
link (or multiloop) invariants and generalized measures on the
space of connections.  In addition, we pose some research problems and
describe some new results, including a proof (due to Sawin) that
the Chern-Simons path integral is not given by a generalized measure.

\end{abstract}

\section{Introduction}

The relation between knots and quantum gravity was discovered in the
course of a fascinating series of developments in mathematics and
physics.  In 1984, Jones \cite{Jones} announced the discovery of a new
link invariant, which soon led to a bewildering profusion of
generalizations.  It was clear early on that these new invariants were
intimately related to conformal field theory in 2 dimensions.  Atiyah
\cite{Atiyah0}, however, conjectured that there should be an
intrinsically 3-dimensional definition of these invariants using gauge
theory.  Witten \cite{Witten2} gave a heuristic proof of Atiyah's
conjecture by deriving the Jones polynomial and its generalizations from
Chern-Simons theory.  The basic idea is simply that the vacuum
expectation values of Wilson loops in Chern-Simons theory are link
invariants because of the diffeomorphism-invariance of the theory.  To
calculate these expectation values, however, Witten needed to use the
relation between Chern-Simons theory and a conformal field theory known
as the Wess-Zumino-Witten (or WZW) model.

In parallel to this work, a new approach to quantum gravity was being
developed, initiated by Ashtekar's \cite{Ashtekar} discovery of the `new
variables' for general relativity.  In this approach, the classical
configuration space is a space of connections, and states of the quantum
theory are (roughly speaking) measures on the space of connections which
satisfy certain constraints: the Gauss law, the diffeomorphism
constraint, and the Hamiltonian constraint.  In an effort to find such
states, Rovelli and Smolin \cite{RS} used a `loop representation' in
which one works, not with the measures per se, but with the expectation
values of Wilson loops with respect to these measures.  In these terms,
the diffeomorphism constraint amounts to requiring that the Wilson loop
expectation values are link invariants.  In itself this was not
surprising; the surprise was that knot theory could be applied to
obtain explicit solutions of the Hamiltonian constraint, as well!

Indeed, in Rovelli and Smolin's original paper they gave a heuristic
construction assigning to each isotopy class of unoriented links a
solution of all the constraints of quantum gravity in the loop
representation.  Later, Kodama \cite{Kodama} showed how to obtain
another sort of solution using Chern-Simons theory.  From Witten's work
it is clear that in the loop representation this solution is just the
Jones polynomial --- or more precisely, the closely related Kauffman
bracket invariant \cite{Kauffman}.

At first these developments may appear to be an elaborate series of
coincidences.  Some of the mystery is removed when we note that the
`Chern-Simons state' of quantum gravity is the only state of a simpler
diffeomorphism-invariant theory in 4 dimensions known as $BF$
theory \cite{BT1,Horowitz}.  However, a truly systematic explanation would
require understanding the following `ladder' of field theories as a
unified structure: general relativity and $BF$ theory in dimension 4,
Chern-Simons theory in dimension 3, and the WZW model in dimension 2.
The concept of a ladder of field theories has appeared in other contexts
and appears to be an important one \cite{Atiyah0,TJZW}.  In Section 1,
after an introduction to the `new variables', we review this ladder of
field theories and its relation to the new knot invariants.

In addition to understanding the Chern-Simons state as a bridge between
knot theory and quantum gravity, there is the much larger task of
making the loop representation of quantum gravity into a mathematically
rigorous theory and justifying, if possible, Rovelli and Smolin's
construction of solutions of the constraint equations from link classes.
One key aspect of this task is to understand the precise sense in
which diffeomorphism-invariant measures on the space of connections
correspond to isotopy invariants of links (or more generally, `multiloops').
In Section 2 we review recent work by Ashtekar, Isham, Lewandowski and
the author \cite{AI,AL,Baez2,Baez2.5,Baez3} on this problem.

In what follows we will not concentrate on the loop representation {\it
per se}, as it is already the subject of a number of excellent review
articles \cite{Ashtekar2,Ashtekar2.5,Bruegmann,Loll,Smolin} and books
\cite{Ashtekar3}.

\section{The New Variables and the Dimensional Ladder}

Traditionally, general relativity has been viewed as a theory in
which a metric is the basic field.   In these terms,
the Einstein-Hilbert action with cosmological constant term is given by
\be        S_{EH}(g) = \int_{M} (R\, + 2\Lambda)\,\vol  ,
\label{EHaction} \ee
where $R$ is the Ricci scalar curvature and $\vol$ is the volume
form associated to the metric $g$ on the oriented 4-manifold
$M$. The equation we get by varying $g$ is, of course,
\[           R_{\mu \nu} - {1\over 2}R g_{\mu \nu} - \Lambda
g_{\mu \nu} = 0.\]
Recent advances in quantizing the theory, however, have taken advantage
of the techniques of gauge theory by emphasizing the role of {\it
connections}.  This approach is also what allows one to relate general
relativity to the `ladder' of simpler field theories shown below.

\vskip 2em

\noindent \hskip 2em
 4d:\hskip 4em {\bf General relativity}
\hskip 3em $\rightarrow$ \hskip 3em {\bf $BF$ theory}

\noindent \hskip 2em
\hskip 24em $\downarrow$

\noindent \hskip 2em
3d: \hskip 17em {\bf Chern-Simons theory }

\noindent  \hskip 2em
\hskip 24em $\downarrow$

\noindent \hskip 2em
2d: \hskip 4em {\bf WZW model} \hskip 1em
$\rightarrow$ \hskip 1em {\bf $G/G$ gauged WZW model}

\vskip 1em
{\begin{center}
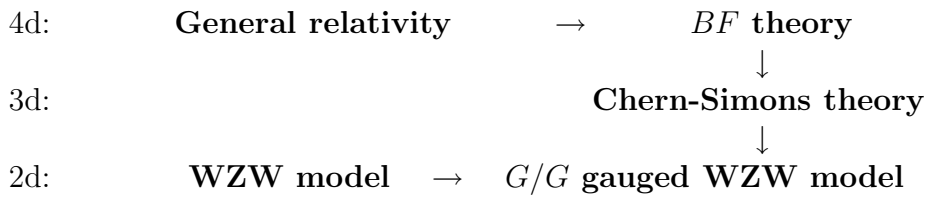
{\tenrm Figure 1.  The dimensional ladder}
\end{center} }\vskip 1.5 em

Historically, the first step towards viewing general relativity as a
gauge theory was the Palatini formalism.  (For a discussion of various
Lagrangians for general relativity, see the review article by Peldan
\cite{Peldan}.)  In this approach, we fix an oriented bundle $\T$ over
$M$ (usually called the `internal space') that is isomorphic to $TM$ and
equipped with a Lorentzian metric $\eta$, and we assume that the
spacetime metric $g$ is obtained from $\eta$ via an isomorphism $e \maps
TM \to \T$.  We may also think of $e$ as a $\T$-valued 1-form, the
`soldering form', and in the Palatini formalism the basic fields are this
soldering form and a connection $A$ on $\T$ preserving the metric
$\eta$, usually called a `Lorentz connection'.  Interestingly, however,
most of what we say below makes sense even when $e \maps TM \to \T$ is
{\it not} an isomorphism.  Thus the Palatini formalism provides a
generalization of general relativity to situations where the metric
$g(v,w) = \eta(e(v), e(w))$ is degenerate.

To clarify the relationship to gauge theory, it is useful to work with
the algebra of differential forms on $M$ taking values in the exterior
algebra bundle $\Lambda \T$.  In particular, the orientation and
internal metric on $\T$ gives rise to an `internal volume form', i.e.\ a
section of $\Lambda^4\T$, and this in turn gives a map from
$\Lambda^4\T$-valued forms to ordinary differential forms, which we
denote by `$\tr$'.  The Palatini action is then given by
\be
S_{Pal}(A,e) = \int_M \tr(e \we e \we F + {\Lambda\over 12} e \we e \we
e \we e)
\label{Palaction} \ee
where we use $\eta$ to regard the curvature $F$ of
$A$ as a $\Lambda^2 \T$-valued 2-form on $M$.  When $A$ corresponds to
the Levi-Civita connection of $g$ via the isomorphism $e \maps TM \to
\T$, the Palatini action equals the Einstein-Hilbert action.  More
importantly, we can obtain Einstein's equations by computing the
variation of the Palatini action.  Using $d_A$ to denote the exterior
covariant derivative of $\Lambda \T$-valued forms, we have:
\ban     \delta S_{Pal}
+ e \we e
&=&  2\int \tr( ((e \we F + {\Lambda \over 6} e \we e \we e) \we \delta e - e
\we d_A e \we \delta A) \ean
where we have ignored boundary terms.  The classical equations of motion
are thus
\[   e \we F + {\Lambda \over 6} e \we e \we e = 0, \quad e \we d_A e =
0.\]
If $e$ is nondegenerate, the latter equation implies that $d_A e
= 0$, i.e., the soldering form is flat, which means that the
connection on $TM$ corresponding to $A$ via the
isomorphism $e$ is torsion-free, hence equal to the Levi-Civita
connection of $g$.  Then the first equation is equivalent to
Einstein's equation (with cosmological constant).

The self-dual formulation of general relativity is based on a slight
variant of the Palatini action, the Plebanski action, that is especially
suited to canonical quantum gravity.  The self-dual formulation applies
very naturally to complex general relativity, and some extra work is
needed to restrict to real-valued metrics.  (In what follows we will
gloss over these very important `reality conditions', on which progress
is just beginning \cite{Ashtekar2.5,ALMMT}.)  The idea is to work with a {\it
complex-valued} soldering form, that is, 1-form on $M$ with values in
the complexified bundle $\C\T$, and a {\it self-dual} connection $A_+$.
To understand this concept of self-duality, note that the internal
metric $\eta$ extends naturally to $\C\T$, making the orthonormal frame
bundle of $\C\T$ into a principal bundle $P$ with structure group
$\SO(4,\C)$.  Now assume we have a spin structure for $\C\T$, that is, a
double cover $\tilde P$ of $P$ with structure group
$\widetilde{SO}(4,\C) = \SL(2,\C) \times \SL(2,\C)$.  Then $\tilde P$ is
the sum $P_+ \oplus P_-$ of `right-handed' and `left-handed' principal
bundles with structure group $\SL(2,\C)$.  This splitting is what lets
us define chiral spinors on $M$.  It is also closely related to duality,
since by using the isomorphism between $\Lambda^2 \C\T$ and $\ad P$ it
lets us write a section $\om$ of $\Lambda^2 \C\T$ as a sum of two parts,
which are precisely the self-dual and anti-self-dual parts with respect
to the `internal' Hodge star operator $\ast$ coming from the internal
metric and orientation on $\C\T$:
\[           \om = \om_+ + \om_-, \qquad  \ast \om_{\pm} = \pm i
\om_{\pm} .\]
We call connections on $P_+$ `self-dual' because, fixing one, we can
identify all the rest with $\Lambda^2 \C\T$-valued 1-forms
that are internally self-dual.  Note that a connection $A$ on $P$ is
equivalent to a pair of connections $A_{\pm}$ on $P_{\pm}$, which we
call its self-dual and anti-self-dual parts.  The curvature $F$ of $A$
is then a $\Lambda^2 \C\T$-valued 2-form which is the sum of the
curvatures $F_\pm$ of $A_\pm$.  Moreover, we have
\[         \ast F_{\pm} = \pm i F_{\pm} .\]

The Plebanski action is then:
\be           S_{Ple}(A_+,e) = \int_M \tr(e \we e \we F_+ +
{\Lambda\over 12} e \we e \we e \we e) \label{Pleaction}. \ee
Just as before, the classical equations of motion are
\[  e \we F_+ + {\Lambda \over 6} e \we e \we e = 0,
\quad e \we d_{A_+} e =
0.\]
To relate these equations to the Palatini formalism, we interpret
$A_+$ as the self-dual part of $A$, so that $F_+$ is the self-dual part
of $F$.  When $e$ is nondegenerate, the second equation
then implies that $A_+$ is the self-dual part of the connection on
$P$ corresponding via $e$ to the Levi-Civita connection on $M$.
The algebraic Bianchi identity then implies that $e \we F_+ = e \we
F$, so the first equation is equivalent to the corresponding equation in
the Palatini formalism, i.e., Einstein's equation.

Now let us turn to $BF$ theory, which is a
diffeomorphism-invariant gauge theory that makes sense in any
dimension.   Suppose that the spacetime manifold $M$ is oriented
and  $n$-dimensional, and that $P$ is a $G$-bundle over $M$, where $G$
is a connected Lie group and
Lie algebra of $G$ is equipped with an invariant
bilinear form which we write as `$\tr$'.   Then the basic fields in
$BF$ theory are a connection $A$ on $P$ and an $(n-2)$-form $B$
with values in $\ad P$, and the action for the theory is
\[       \int_M \tr(B \we F)  .\]
After attention was drawn to it by the work of Blau and Thompson
\cite{BT1} and Horowitz \cite{Horowitz}, this theory has been
extensively studied in dimensions 2, 3, and 4.  In dimension 2, it is
closely related to Yang-Mills theory \cite{Witten3}.  In dimension 3, it
has gravity in the Palatini formalism as a special case
\cite{AHRSS,Witten1}.  In dimension 4, it is also known as `topological
gravity' when we take $G = \SL(2,\C)$ and take $P$ to be the bundle
$P_+$ used in the self-dual formulation of general relativity
\cite{CM,CS}.  Mathematically, $BF$ theory is closely related to moduli
spaces of flat connections, and thereby to the Ray-Singer torsion, the
Alexander-Conway polynomial invariant of links, and the Casson invariant
of homology 3-spheres \cite{BBRT,BT1,CCM,Schwartz}.

In what follows we will focus on dimension 4, and consider
a variant of the $BF$ action that includes a $B \we B$ term:
\be   S_{BF}(A,B)  = \int_M \tr (B \we F + {\Lambda \over 12} B \we B) .
\label{BFaction} \ee
Ignoring boundary terms, the variation of the action is then
\ban     \delta S_{BF}
&=&
\int \tr ((F + {\Lambda \over 6} B) \we \delta B - d_A B \we \delta A) ,
\ean
where $d_A B$ denotes the
exterior covariant derivative of $B$.  Setting $\delta S_{BF} = 0$
we obtain the classical equations of motion:
\[         F + {\Lambda\over 6} B = 0, \quad d_A B= 0 .\]
Note that the case $\Lambda \ne 0$ is very different from the case
$\Lambda = 0$.  When $\Lambda \ne 0$, the second equation follows from
the first one and the Bianchi identity, so $A$ is {\it arbitrary} and it
determines $B$.  When $\Lambda = 0$, $A$ must be {\it flat} and $B$ is
any section with $d_A B = 0$.

Note that $BF$ theory is very similar to general relativity in its
self-dual formulation, with $B$ playing the role of $e \we e$.  To
compare these theories more precisely, we will write simply $P$ for the
`right-handed' $\SL(2,\C)$ principal bundle $P_+$ discussed above, and
drop the subscript `$+$' on the $A$ and $F$ fields.  Now, there is a
mapping from the space of fields $(A,e)$ for general relativity to the
space of fields $(A,B)$ for $BF$ theory (with $G = \SL(2,\C)$) given by
\[          (A,e) \mapsto (A,B) = (A,e \we e) .\]
If $(A,e\we e)$ is a solution of the $BF$ equations of motion, then
$(A,e)$ is a solution of Einstein's equations.  Of course, we obtain
only a limited class of solutions of Einstein's equations this way: for
$\Lambda = 0$ we obtain precisely the flat solutions, while for $\Lambda
\ne e$ we obtain those with $F = -{\Lambda \over 6} e \we e$.

Amazingly, $BF$ theory appears to yield solutions of the constraint
equations of {\it quantum} gravity by a similar mechanism.  Moreover,
these solutions are closely related to well-known link invariants.  No
formalism for quantum gravity has been worked out to the point where we
can feel full confidence in these results, but the work of various
authors using the connection \cite{Kodama} and loop \cite{BGP}
representations, as well as the BRST formalism \cite{CS}, all seems to
point in the same direction.  In what follows we will describe these
results in terms of Dirac's approach to canonical quantization of
constrained systems.

Suppose, then, that $M = \R \times S$, and identify $S$ with the slice
$\{t = 0\}$.  Working in temporal gauge, both classical $BF$ theory and
classical general relativity in the Ashtekar formalism can be described
in terms of a `kinematical' phase space $T^\ast \A$ together with
certain constraints.  Here the configuration space $\A$ consists of
connections on the bundle $P|_S$.  A tangent vector $\delta A \in T_A
\A$ can be identified with an $\ad P$-valued 1-form on $S$, so a
cotangent vector can be identified with an $\ad P$-valued 2-form $B$,
using the pairing
\[       \langle B, \delta A\rangle = \int_S  \tr(B \we \delta A) .\]

In $BF$ theory, one obtains a point in the kinematical phase space from
a solution of the equations of motion by restricting $A$ and $B$ to $S$,
while in general relativity one does the same with $A$ and the self-dual
part of $e \we e$, regarded as an $\ad P$-valued 2-form.  In general
relativity, however, it is conventional to use the isomorphism
\[
\Lambda^2 T^\ast M \cong TM \tensor \Lambda^3 T^\ast M \]
to think of $(e \we e)_+$ as an $\ad P$-valued `vector density', usually
written $\tilde E$.  This is precisely where the advantage of the
self-dual formalism over the Palatini one appears: one can attempt a
similar trick in the Palatini formalism, but in that case, extra
constraints negate the advantage of working with this formalism.  In the
self-dual formalism, no conditions on $\tilde E$ need hold for it to
come from a complex soldering form $e$.

Since the kinematical phase space for $BF$ theory is the same as that
for general relativity, the difference between the theories lies in the
constraints.  To describe these, it is handy to introduce indices
$i,j,k,\dots$ labeling a basis of sections of $TS$, and indices
$a,b,c,\dots$ labeling an orthonormal basis of sections of $\ad P|_S$.
(Note that $TS$ and $P|_S$ are trivial so we can find global bases of
sections.)  In $BF$ theory the canonically conjugate variables can then
be written as $A_i^a$ and a vector density $\tilde B^i_a = {1\over
2}\epsilon^{ijk} B_{jka}$, and the constraints are the Gauss law
\[       \G^a = \partial_i \tilde B^{ia} + [A_i, \tilde B^i]^a = 0\]
 and
\[      \Con^a_{ij} =
F_{ij}^a + {\Lambda \over 6} \epsilon_{ijk} \tilde B^{ka} = 0. \]
In general relativity, on the other hand, working with the lapse and
shift as Lagrange multipliers, the canonically conjugate
variables are $A_i^a$ and $\tilde E^i_a$, and the constraints are
the Gauss law
\[    \G^a = \partial_i \tilde E^{ia} + [A_i,\tilde E^i]^a ,\]
together with the Hamiltonian and diffeomorphism constraints:
\[      \H = \epsilon_{abc} \tilde
E^{ia} \tilde E^{jb} F_{ij}^c + {\Lambda \over 6}
\epsilon_{abc}\epsilon_{ijk}  \tilde E^{ia} \tilde E^{jb} \tilde
E^{jc} , \qquad        \H_j = \tilde E^i_a F^a_{ij}   .\]

We now turn to the canonical quantization of general relativity and $BF$
theory.  (We emphasize that the remarks above could be made rigorous in
a rather straightforward way, while the rest of this section is
heuristic in character.)  In either theory, we begin with a `kinematical
state space' consisting of functions on $\A$, the space of connections
on $P|_S$.  In $BF$ theory, we then quantize the canonically conjugate
fields $A_i^a$ and $\tilde B^i_a$, making them into operators on the
kinematical state space by
\[        (\hat A_i^a(x) \psi)(A) = A_i^a(x)\psi(A), \quad
          (\hat {\tilde B}{}^i_a(x) \psi)(A) = {\delta \psi \over \delta
A_i^a(x)}(A).\]
For a function on $\A$ to represent a physical state,
it must be annihilated by the quantized constraints:
\[           \hat \G^a \psi = 0, \quad \hat \Con^a_{ij} \psi = 0.\]
Classically the Gauss law generates gauge transformations, so the
best interpretation of the first equation is simply that $\psi$ is
invariant under small gauge transformations.  The second equation is a
first-order partial differential equation on $\A$:
\[      {\Lambda \over 6} \epsilon_{ijk}{\delta \over \delta
A_{ka}}\psi = - F_{ij}^a \psi .\]
For $\Lambda \ne 0$ this has a single solution, the so-called
`Chern-Simons state':
\[        \psi_{CS}(A) = e^{-{6\over \Lambda}\int_S \tr(A \we
dA + {2\over 3} A \we A \we A) }  ,\]
which is automatically invariant under small gauge transformations.  For
$\Lambda = 0$, any gauge-invariant $\psi$ supported on the space of {\it
flat} connections is a solution.  We call such solutions `flat states'.
For $\Lambda = 0$ we can also use the `constrain before quantizing'
strategy and describe the flat states as functions on the moduli space
of flat connections on $P|_S$, which has the advantage over $\A$ of
being finite-dimensional.  (For more on the flat states, see the work of
Blencowe \cite{Blencowe}.)

One can attempt to quantize gravity in a similar fashion,
defining operators on ${\bf H}_{kin}$ by
\[     (\hat A_i^a(x) \psi)(A) = A_i^a(x)\psi(A), \quad
      (\hat {\tilde E}{}^i_a(x) \psi)(A) = {\delta \psi \over \delta
A_i^a(x)}(A),\]
and seeking solutions of the constraint equations:
\[        \hat \G^a \psi = \hat \H_j \psi = \hat H \psi = 0 .\]
The remarkable thing is that the solutions we found for $BF$ theory
are also annihilated by these constraints,
at least if we take the operator ordering for $\hat \H$ given by
\[   \hat \H = \epsilon^{abc} {\delta  \over \delta
A_i^a} {\delta  \over \delta A_j^b}
F_{ijc} + {\Lambda \over 6}\epsilon^{abc}\epsilon_{ijk}
{\delta  \over \delta A_i^a} {\delta  \over \delta
A_j^b} {\delta  \over \delta A_k^c}  .\]
The reason is simple.  In classical general relativity the constraint
$\G^a$ generates gauge transformations, while $\H_j$ generates
diffeomorphisms, so we should interpret the quantized constraint
equations $\hat G^a \psi = \hat \H_j \psi = 0$ as saying that $\psi$ is
invariant under small gauge transformations and diffeomorphisms.  A
mathematically more proper way to state this is to say that $\psi$ is
invariant under small automorphisms of the bundle $P|_S$.  (It is
unclear whether one should also demand invariance under `large' bundle
automorphisms; we will not do so here.)  The solutions we found for $BF$
theory are indeed invariant under small bundle automorphisms!  The
Hamiltonian constraint, on the other hand, can be expressed in terms of
the constraints of $BF$ theory by
\[        \hat \H =  \epsilon^{abc} {\partial\over
\partial A_{ia}}  {\partial\over \partial A_{jb}}
 \hat \Con_{ijc} ,\]
if we identify $\tilde E^{ia}$ in general relativity with $\tilde
B^{ia}$ in $BF$ theory.  Thus $\hat \Con_{ij}^a\psi = 0 $ implies $\hat
\H\psi = 0$.

In short, the relationship between $BF$ theory and general relativity
gives us some {\it explicit} solutions of the constraint equations of
quantum gravity: the Chern-Simons state when $\Lambda \ne 0$, and the
flat states when $\Lambda = 0$.  What is the physical significance of
these solutions?  As noted by Kodama \cite{Kodama}, if $S = S^3$, the
Chern-Simons state appears to represent a `quantized deSitter universe'
(or anti-deSitter, depending on the sign of $\Lambda$).  Smolin and Soo
have recently done some fascinating work on the `problem of time' using
this idea \cite{SS}.  Similarly, if $S = \R^3$ it appears that the
single flat state represents a `quantized Minkowski space'!  However,
there has been some debate over whether the Chern-Simons state is
normalizable, and the same could be asked of the flat states.  We will
have more to say about this question in the next section, but it can
only really be settled when we understand the problem of the inner
product in quantum gravity.  It is worth noting here that some
approaches to the inner product problem rely heavily on ideas from knot
theory and 3-dimensional topology.  For example, Rovelli has drawn
inspiration from the Turaev-Viro theory, a topological quantum field
theory in 3 dimensions, to give a formula for the physical inner product
\cite{Rovelli2}, which unfortunately is purely formal at present.  An
alternative strategy, which is mathematically rigorous but physically
more radical, is to split $S$ into two manifolds with boundary, and to
use the Chern-Simons state on $S$ to define inner products of `relative
states' on each of the two halves \cite{Baez3,Crane}.

Now let us return to the Chern-Simons state of $BF$ theory with
arbitrary gauge group $G$.  An interesting relation to knot theory shows
up when we try to compute the `loop transform' of this state.  Given
loops $\gamma_i$ in $S$, the loop transform of $\psi_{CS}$ is formally
given by
\[     \hat \psi_{CS}(\gamma_1, \dots, \gamma_n) =
\int_{\A} \,\prod_{i=1}^n \tr(T e^{\int_{\gamma_i} A}) \, \psi_{CS}(A)\,
{\cal D}A, \]
where ${\cal D}A$ is purely formal `Lebesgue measure' on $\A$.  Since
$\psi_{CS}$ is invariant under small bundle automorphisms, if we assume
${\cal D}A$ shares this invariance property we can conclude that $\hat
\psi_{CS}$ is a `multiloop invariant', that is, it should not change
when we apply a given small diffeomorphism to all of the loops
$\gamma_i$.  In particular, if we restrict to {\it links} (embedded
collections of loops), $\hat \psi_{CS}$ should give a link invariant.

This reasoning is merely heuristic, due to the mysterious nature of
`${\cal D}A$', but in fact Witten \cite{Witten2} was able to compute the
link invariant corresponding to $\hat \psi_{CS}$ for $G = \SU(n)$, and
similar computations are now possible for many other groups.  For
$\SU(2)$ the result is simply the Kauffman bracket, which is a link
invariant defined by the skein relations shown below, and normalized so
that its value on the empty link is 1.

\vskip 1.5 em

\begin{center}
\setlength{\unitlength}{0.0125in}%
\begin{picture}(160,40)(90,700)
\thicklines
\put( 30,740){\line(-1,-2){ 20}}
\put( 30,700){\line(-1, 2){ 7}}
\put( 10,740){\line( 1,-2){ 7}}
\put(90,740){\line( 0,-1){  40}}
\put(110,740){\line( 0,-1){  40}}
\put(170,740){\oval( 20, 20)[bl]}
\put(170,740){\oval( 20, 20)[br]}
\put(170,700){\oval( 20, 20)[tr]}
\put(170,700){\oval( 20, 20)[tl]}
\put( 45,715){\makebox(0,0)[lb]{\raisebox{0pt}[0pt][0pt]{\twlrm $= \;\;
q^{1\over 4}$}}}
\put(120,715){\makebox(0,0)[lb]{\raisebox{0pt}[0pt][0pt]{\twlrm$\; +\;
q^{-{1\over 4}}$}}}

\put(240,720){\circle{22}}
\put(270,715){\makebox(0,0)[lb]{\raisebox{0pt}[0pt][0pt]{\twlrm$ =
-(q^{1\over 2} + q^{-{1\over 2}})$}}}

\end{picture}
\end{center}

\vskip 1em
{\begin{center}
{\tenrm Figure 2. Skein relations for the Kauffman bracket}
\end{center} }\vskip 1.5em

\noindent Here
\[         q = e^{2\pi i/(k + 2)}  , \qquad k = {24\pi i\over \Lambda}
.\]
Only for integer $k$ is $\psi_{CS}$ invariant under
large gauge transformations.  It is important to note that the Kauffman
bracket is an invariant of {\it framed} links, reflecting the fact that
one must regularize the Wilson loops to properly calculate their
expectation value

For the groups $\SU(n)$ one obtains a link invariant generalizing the
Kauffman bracket known as the HOMFLY polynomial, while for $\SO(n)$ one
obtains yet another invariant, the Kauffman polynomial
\cite{Kauffman,RT}.  All of these link invariants are also defined by
skein relations.  Since $\psi_{CS}$ may be defined as the unique
function on $\A$ annihilated by the constraint $\hat \Con^a_{ij}$, while
$\hat \psi_{CS}$ is determined (at least on links) by the skein
relations, it appears that the skein relations are simply a rewriting of
the constraint $\hat \Con^a_{ij}$ in the language of Wilson loops.
The author has speculated on the implications of this idea for physics
elsewhere \cite{Baez4}, but it also suggests the following essentially
mathematical problem:

\begin{problem} \et Derive the skein relations for the Kauffman and
HOMFLY polynomials as directly as possible from the corresponding
$\SU(n)$ and $\SO(n)$ $BF$ theories in 4 dimensions.  (Hint: study the
existing work on deriving the skein relations via loop deformations
\cite{Bruegmann2}, and consider the possibility of a relation to the
theory of surfaces immersed in 4-manifolds \cite{CM}.)  \end{problem}

It is often tacitly assumed that the Chern-Simons state for $\SL(2,\C)$,
which is the one relevant to quantum gravity, has the Kauffman bracket
as its loop transform just as the $\SU(2)$ Chern-Simons state does.
Perturbative calculations on $S^3$ appear to support this
\cite{BarNatan2}, but nonperturbatively, especially on 3-manifolds with
nontrivial fundamental group, the situation is far from clear:

\begin{problem} \et  Describe Chern-Simons theory with noncompact
gauge group (in particular, $\SL(2,\C)$) as a topological
quantum field theory satisfying axioms similar to those listed by Atiyah
\cite{Atiyah}, and compute the vacuum expectation values of
Wilson loops in this theory.  (Hint: see the work of Bar-Natan and Witten
\cite{BW,Witten4}.)  \end{problem}

Let us now turn to how one computes the link invariant
corresponding to $\hat \psi_{CS}$.  For reasons of space we will be
very sketchy here.  First, writing
\[      \psi_{CS}(A) = e^{-{6\over\Lambda} S_{CS}(A)}  ,\]
the quantity
\be      S_{CS}(A) = \int_S \tr(A \we dA + {2\over 3} A \we A \we A)
\ee
can be interpreted as the action for a 3-dimensional field theory,
Chern-Simons theory.  Note that the 3-manifold $S$, which played the
role of `space' in $BF$ theory, now plays the role of `spacetime', and
that the loop transform of $\psi_{CS}$ can now be thought of as a path
integral.  To compute this path integral, one chops up $S$ (and the link
in $S$) into simple pieces, deals with these pieces, and then glues them
together using the axioms of a topological quantum field theory.  In
particular, it is useful to begin by considering Chern-Simons theory on
a spacetime $S = \R \times \Sigma$, with $\Sigma$ a Riemann surface.
This lets us descend the dimensional ladder yet another rung, since in
this situation the states of Chern-Simons theory correspond exactly to
the conformal blocks of the WZW model \cite{Witten2}, and we can derive
the Kauffman bracket skein relations from the transformation properties
of $n$-point functions under elements of the mapping class group.
However, more recently it has become clear that the more fundamental
relation is that between Chern-Simons theory and a 2-dimensional {\it
topological} quantum field theory, the $G/G$ gauged WZW model
\cite{BT2}.  From this point of view, the WZW model itself serves mainly
as a computational tool.

What is the real meaning of the dimensional ladder?  Most importantly,
one climbs down it by considering `boundary values'.  For example, on
4-manifolds without boundary, when $\Lambda$ is nonzero $S_{BF}$ is
unchanged by infinitesimal transformations of the form
\[           A \mapsto A + \delta A, \qquad B \mapsto B - {6\over
\Lambda} d_A \delta A , \]
since such transformations change the Lagrangian
by an exact form:
\ban      \delta \tr(B \we F + {\Lambda\over 12}B \we B)
&=& -{6\over \Lambda} d\, \tr(\delta A \we F) \ean
This symmetry is the reason why any connection $A$ gives a solution of
the classical equations of motion.  On a 4-manifold with boundary, the
exact form gives a boundary term.  Up to a constant factor, this is
precisely the variation of the Chern-Simons action:
\[ \delta S_{CS}(A)
= 2 \int_S \tr(\delta A \we F) .\]
(It is the relation between the
$BF$ Lagrangian and the 2nd Chern form $\tr(F \we F)$ that makes this
computation work \cite{TJZW}.)  In a similar but subtler manner,
$S_{CS}(A)$ is invariant under small gauge transformations when the
3-manifold $S$ has no boundary, but by considering how it changes when
$S$ has boundary we may derive the action of the $G/G$ gauged WZW model
\cite{BT2}.

\begin{problem} \et Give, if possible, a construction of $BF$ theory in
4 dimensions as a topological quantum field theory.  (In the $\Lambda =
0$ case, the Atiyah axioms \cite{Atiyah} will need to be generalized to
treat situations where the Hilbert space of states is
infinite-dimensional.)  \end{problem}

We conclude this section with a few words about the $\Lambda \to 0$
limit of the Chern-Simons state and Vassiliev invariants.  If we regard
$\psi_{CS}$ as a function of $\Lambda$, the $\Lambda \to 0$ limit
appears to be very singular.  And indeed, we have seen that the
character of $BF$ theory becomes very different when $\Lambda$ vanishes.
However, when we consider {\it imaginary} $\Lambda$, corresponding to
integer $k$, the formula for $\hat \psi_{CS}$ becomes an oscillatory
integral which can be approximated as $\Lambda \to 0$ using the method
of stationary phase \cite{Witten2}.  The points of stationary phase are,
by the above formula for $\delta S_{CS}$, precisely the {\it flat}
connections.  Thus we expect that as $\Lambda \to 0$, the Chern-Simons
state approaches a particular flat state.  Indeed, the HOMFLY and
Kauffman polynomials can all be expanded as power series in $\Lambda$,
with coefficients being link invariants of a special sort known as
invariants of finite type, or Vassiliev invariants \cite{BarNatan}.  If
we accept the assumption that the Chern-Simons state for $\SL(2,\C)$
corresponds to the Kauffman bracket, at least on $S^3$, we obtain a
fascinating relation between quantum gravity and Vassiliev invariants.
For more on this, we urge the reader to the references
\cite{Baez0,BGP,GP,Kauffman2}.

\section{Multiloop Invariants and Generalized Measures}

In the previous section, much of the discussion of {\it quantum} gravity
in 4 dimensions was heuristic in character.  In particular, we imagined
starting with a kinematical state space ${\bf H}_{kin}$ consisting of
functions on the space of connections $\A$, and defining the physical
state space ${\bf H}_{phys}$ to consist of those $\psi \in {\bf
H}_{kin}$ satisfying the Gauss law, diffeomorphism constraint, and
Hamiltonian constraint.  To make this rigorous, we should try to
give a precise definition ${\bf H}_{kin}$, and then make sense of the
constraints.

As already noted, the Gauss law and diffeomorphism constraints have such
a simple geometrical meaning that we can make sense of them quite nicely
without defining operators corresponding to these constraints, or even
choosing a specific definition of ${\bf H}_{kin}$.  Namely, we can take
these constraints to say that $\psi$ is invariant under small
automorphisms of the bundle $P|_{S}$.  It is far more difficult to treat
the Hamiltonian constraint properly.  In what follows we will discuss a
particular way of definining ${\bf H}_{kin}$, first suggested by Rovelli
and Smolin under the name of the `loop representation' \cite{RS}, and
subsequently made rigorous by Ashtekar, Isham, Lewandowski and the
author \cite{AI,AL,Baez2,Baez2.5,Baez3}.  According to the original
heuristic work of Rovelli and Smolin \cite{RS}, a large space of
solutions of the Hamiltonian constraint can be described explicitly
using the loop representation!  However, finding a rigorous formulation
of the Hamiltonian constraint in the loop representation of quantum
gravity remains one of the outstanding challenges of the subject.

At the heuristic level, the
key ingredient of the loop representation is the loop transform
\[      \hat\psi(\gamma_1, \dots, \gamma_n) =
\int_{\A} \,\prod_{i=1}^n \tr(T e^{\int_{\gamma_i} A}) \, \psi(A)\,
{\cal D}A\]
taking functions on the space of connections to functions of multiloops.
Unfortunately, the `Lebesgue measure' ${\cal D}A$ is a purely formal
object!  There is, however, a way to avoid this problem.  The idea is to
treat states in ${\bf H}_{kin}$ not as {\it functions} on the space of
connections, but as `generalized measures' on the space of connections.
This amounts to treating the combination $\psi(A) {\cal D}A$ as a single
object, to be made sense of in its own right.  As we shall see, there is
a way to do this which gives us access to a large class of $\psi \in
\H_{kin}$ that are invariant under small bundle automorphisms.  Since
much of what follows is applicable to any smooth manifold $M$ and
principal $G$-bundle $P$ over $M$, where $G$ is a compact connected Lie
group, we will work at this level of generality, and let $\A$ denote the
space of smooth connections on $P$.

Working with measures on an infinite-dimensional space like $\A$
is a notoriously tricky business, but if we take the attitude
that the job of a measure is to let us integrate functions,  we
can simply specify an algebra of functions on $\A$ that we would
like to integrate, and define generalized measures on $\A$ to be
continuous linear functionals on this algebra.  In the case at
hand we want this algebra to contain the Wilson loop functions
$\tr( T e^{\int_\gamma A})$.  So, suppose that $\gamma$ is
a piecewise smooth path in $M$, and let $\A_\gamma$ denote the
space of smooth maps $F\maps P_{\gamma(a)} \to P_{\gamma(b)}$ that
are compatible with the right action of $G$ on $P$:
\[            F(xg) = F(x)g .\]
Note that for any connection $A \in \A$, the parallel transport
map
\[        T e^{\int_\gamma A}: P_{\gamma(a)} \to P_{\gamma(b)} \]
lies in $\A_\gamma$.  Of course, if we fix a
trivialization of $P$ at the endpoints of $\gamma$,
we can identify $\A_\gamma$ with the group $G$.    Now let
$\Fun_0(\A)$ be the algebra of functions on $\A$ generated by
those of the form
\[            f(T e^{\int_\gamma A})  \]
where $f$ is a continuous function on $\A_\gamma$.  Let $\Fun(\A)$
denote the completion of $\Fun_0(\A)$ in the sup norm:
\[          \|\psi \|_\infty = \sup_{A \in \A} |\psi(A)|. \]
It is easy to check that the Wilson loops lie in this algebra,
taking the trace in any finite-dimensional representation of $G$.
Thus we define the space of `generalized measures' on $\A$
to be the dual $\Fun(\A)^\ast$.  Given a function $\psi \in \Fun(\A)$, we
can write $\mu(\psi)$ as an integral
\[          \int_{\A} \psi(A) \, d\mu(A) \]
if we wish to emphasize that $\mu$ serves the same purpose as a measure
on $\A$.

We can now make the relation between knot theory and
diffeomorphism-invariant gauge theory precise, as follows.  For
simplicity we consider only the case $G = \SU(n)$.
Suppose $\mu$ is a generalized measure on $\A$ that is invariant under
all small bundle automorphisms.  Then the quantity
\[  \hat \mu(\gamma_1, \dots, \gamma_n) =  \int_\A \,\prod_{i=1}^n \tr(T
e^{\int_{\gamma_i} A} \, d\mu(A), \]
where we take traces in the fundamental representation,
is a multiloop invariant.   Conversely, knowing the multiloop invariant
$\hat \mu$ determines $\mu$ uniquely!  A basic problem is:

\begin{problem}\et  Characterize the multiloop invariants
that arise from generalized measures on the
space $\A$ of smooth connections on a given bundle $P$.  \end{problem}

\noindent It is worth emphasizing that while $\hat \mu$ restricts to a
link invariant, the link invariant is not enough to determine $\hat\mu$.
The point is that generalized measures on $\A$ can give multiloop
invariants that detect singularities: self-intersections, corners, cusps
and the like \cite{Baez2}.  This may be a good thing for quantum
gravity, since the Hamiltonian constraint is also sensitive to
self-intersections \cite{RS}.  Also, within knot theory itself, more and
more attention is being paid to multiloops with self-intersections
\cite{BarNatan,Birman}.

It is typical that the measures appearing in quantum field theory
(either as path-integrals or as states in the canonical formalism) are
not supported on the space of smooth fields \cite{MM}, but on a larger
space of `distributional fields'.  And indeed, generalized measures on
$\A$ can alternatively be described as honest measures on a space
$\overline \A$ of `generalized connections' containing $\A$ as a dense
subset.  These `generalized connections' are objects that allow parallel
transport along paths in $M$, but without some of the smoothness
conditions characteristic of connections in $\A$.  In particular, like a
connection, a generalized connection $A$ associates to each piecewise
smooth path $\gamma\maps [a,b] \to M$ an element of $\A_\gamma$, which
we write formally as $Te^{\int_\gamma A}$.  Generalized connections are
only required to satisfy the following property: given any finite set of
piecewise smooth paths $\gamma_i$, $1 \le i \le n$, and any continuous
function $F \maps \prod A_{\gamma_i} \to \C$, if
\[         F(Te^{\int_{\gamma_1} A}, \dots, Te^{\int_{\gamma_n} A}) = 0
\]
for all smooth connections $A \in \A$, then the same equation holds for
any generalized connection.  (It follows that parallel transport using
generalized connections composes when one composes paths, is independent
of the parametrization of the path, and so on.)  We equip $\overline \A$
with the weakest topology such that all the maps
\[               A \mapsto T e^{\int_\gamma A}  \]
are continuous from $\overline \A$ to $\A_\gamma$ (identifying the
latter space with $G$).  In this topology $\overline \A$ is a compact
Hausdorff space, and $\A$ is dense in $\overline \A$.  Finally,
generalized measures on $\A$ are in one-to-one correspondence with
measures on $\overline \A$.   (Here and in what follows by `measure' we
implicitly mean `finite regular Borel measure.')

Now, it is common in quantum field theory to avoid certain infinities in
integrals over $\A$ by integrating instead over $\A/\G$, where $\G$ is
the gauge group of $P$.  It is worth digressing for a moment to explain
why this is unnecessary here.  While much work on the loop
representation uses generalized measures on $\A/\G$, these are in
one-to-one correspondence with gauge-invariant generalized measures on
$\A$, as follows.  First, define the `holonomy C*-algebra' $\Fun(\A/\G)$
to be the algebra of functions on $\A/\G$ that pull back under the
quotient map $p\maps \A \to \A/\G$ to elements of $\Fun(\A)$.  The
functions $p^\ast \psi \in \Fun(\A)$ one obtains this way are precisely the
gauge-invariant elements of $\Fun(\A)$.    Next, define a generalized
measure on $\A/\G$ to be an element of the dual $\Fun(\A/\G)^\ast$.
Then given a gauge-invariant generalized measure $\tilde \mu$ on $\A$,
there is a generalized measure $\mu$ on $\A/\G$ given by
\[         \mu(\psi) = \tilde\mu(p^\ast \psi)  \]
for all $\psi \in \Fun(\A/\G)$.  Conversely, given a generalized measure
$\mu$ on $\A/\G$, we can obtain the corresponding gauge-invariant
generalized measure $\tilde \mu$ on $\A$ as follows.  There is a
rigorous way to average over the gauge group action, giving a continuous
linear map $q \maps \Fun(\A) \to \Fun(\A/\G)$ with the property that
$qp^\ast$ is the identity on $\Fun(\A/\G)$.  Given $\mu$, we then define
$\tilde \mu$ by
\[         \tilde \mu(\psi) = \mu(q\psi)  \]
for all $\psi \in \Fun(\A)$.  (These results follow from a slight extension
of published work \cite{Baez3}.)

Almost everything one does with measures can be done with generalized
measures.  This should not be surprising, since generalized measures on
$\A$ {\it are} measures on $\overline \A$.  However, it is rarely
necessary to refer to the big space $\overline \A$.  For example, a
generalized measure $\mu$ is said to be `strictly positive' if $\psi \ge 0$
and $\psi \ne 0$ implies $\mu(\psi) > 0$.  Given a strictly positive
generalized measure $\tilde \mu$ on $\A$, we can form a Hilbert space
$L^2(\A,\mu)$ by completing $\Fun(\A)$ in the norm
\[      \|\psi\|_2 = \mu(\overline \psi \psi)^{1/2}   .\]
If in addition $\tilde \mu$ is invariant under some group, then this
group will have a unitary representation on $L^2(\A,\tilde\mu)$.  In
particular, if $\tilde\mu$ is gauge-invariant, the corresponding
generalized measure $\mu$ on $\A/\G$ will also be strictly positive,
allowing us to form $L^2(\A/\G,\mu)$ in a similar fashion, and
$L^2(\A/\G,\mu)$ will be isomorphic as a Hilbert space to the subspace
of gauge-invariant elements of $L^2(\A, \tilde \mu)$.

How does one {\it construct} generalized measures on $\A$, however?
Without a way to do this, the theory would be of little interest.  There
are various ways; unfortunately, most of them currently require one to
work with piecewise analytic paths rather than piecewise smooth paths as
we have done so far.  (The reason is that piecewise smooth paths can
have horribly complicated self-intersections.)  Everything we have said
so far about the loop representation is still true if we assume that $M$
is real-analytic and all paths are piecewise analytic; the bundle $P$
and the connections in $\A$ can still be merely smooth.  Henceforth we
will assume this is the case, and define $\Diff(M)$ to consist of {\it
analytic} diffeomorphisms of $M$, and $\Aut(P)$ to consist of bundle
automorphisms that act on the base space $M$ by analytic
diffeomorphisms.

\begin{problem} \et  Determine which of the results below
can be generalized to the smooth category.    \end{problem}

The most basic recipe for constructing generalized measures is a
nonlinear version of the theory of `cylinder measures' widely used to
study linear quantum fields.  Interestingly, this recipe is based on
ideas from lattice gauge theory.  In lattice gauge theory one
approximates the space of connections on $\R^n$ by the space of
connections on a lattice in $\R^n$, where a connection on the lattice
assigns a group element to each edge of the lattice.  In the present
diffeomorphism-invariant context we must consider all graphs embedded in
the manifold $M$.  An `embedded graph' $\phi$ in $M$ is a collection
analytic paths $\phi_i \maps [0,1] \to M$, $1 \le i \le n$, called
`edges', such that

\vskip 1em
1. for all $i$, $\phi_i$ is one-to-one,

2. for all $i$, $\phi_i$ is an embedding when restricted to $(0,1)$,

3.  for all $i \ne j$, $\phi_i[0,1]$ and $\phi_j[0,1]$ intersect, if
at all, only at their endpoints.

\vskip 1em \noindent
Given an embedded graph $\phi$, we define the space $\A_\phi$ of
connections on $\phi$ as follows:
\[        \A_\phi = \prod_{i = 1}^n \A_{\phi_i}  .\]
If we trivialize $P$ at the endpoints of the edges of $\phi$, we can
identify $\A_\phi$ with a product of copies of $G$.

Now, given embedded graphs $\phi$ and $\psi$, let us write $\phi
\hookrightarrow \psi$ if every edge of $\phi$ is, up to
reparametrization, a product of edges of $\psi$ and their inverses.  If
$\phi \hookrightarrow \psi$, there is a natural map from $\A_\psi$ onto
$\A_\phi$.  We say that a family of measures $\{\mu_\phi\}$ on the
spaces $\A_\phi$ is `consistent' if whenever $\phi \hookrightarrow
\psi$, the measure $\mu_\psi$ pushes forward to the measure $\mu_\phi$
under this natural map.  Every generalized measure on $\A$ uniquely
determines a consistent family of measures $\{\mu_\phi\}$.  Conversely
--- and this is how one can construct generalized measures --- every
consistent family of measures $\{\mu_\phi\}$ that is uniformly bounded
in the usual norm determines a unique generalized measure
on $\A$.

For example, if one uses products of copies of the normalized Haar
measure on $G$ to define measures on the spaces $\A_\phi$, one can
easily check the consistency and boundedness conditions and obtain a
generalized measure $\tilde\mu$ on $\A$ called the `uniform' generalized
measure \cite{AL,Baez3}.  This is a kind of partial substitute for the
nonexistent Lebesgue measure ${\cal D}A$.  In particular, $\tilde \mu$
is strictly positive and invariant under all automorphisms of the bundle
$P$.  It therefore gives rise to a generalized measure on $\A/\G$,
the `Ashtekar-Lewandowski' generalized measure.


Moreover, given a function $\psi \in \Fun(\A)$ that is invariant
under small bundle automorphisms, the product $\psi \tilde \mu$ will be
a generalized measure invariant under small automorphisms, or in other
words, a solution to the Gauss law and diffeomorphism constraints.  We
do not expect to find many solutions this way, though; there are simply
not enough functions $\psi$ with this property.  However, a more
sophisticated version of this approach {\it does} give many solutions,
indeed, one for each isotopy class of links \cite{Ashtekar2.5}!  It
would be very interesting to know whether, as in the heuristic work of
Rovelli and Smolin, these are also solutions of the Hamiltonian
constraint.

One can use the same basic recipe to construct other
generalized measures on $\A$ that are invariant under small bundle
automorphisms.  Examples include those whose loop transforms are
multiloop invariants detecting singularities \cite{Baez2}.

Let us conclude by returning to the exact solutions we discussed in the
previous section: the Chern-Simons state and flat states.  Are these
given by generalized measures on the space of connections?  In the case
of the flat states the answer is yes: it is easy to see that every
measure on the moduli space of flat connections on $P$ gives a
generalized measure on $\A$ that is invariant under small bundle
automorphisms \cite{Baez2}.

For the Chern-Simons state the question is not quite well-posed as it
stands!  The problem is that the Kauffman bracket of a link depends on a
framing, while the multiloop invariants coming from generalized measures
do not.  One strategy to deal with framing issues is to work with an
algebra generated by regularized Wilson loop observables, such as the
`tube algebra' \cite{Baez2.5}.  This leads to an alternate definition
of generalized measure, and such generalized measures determine
framed link invariants.  However, the following argument due to
Sawin \cite{Sawin} shows that the Kauffman bracket cannot come from such
a generalized measure, at least not for $q$ a root of unity near 1.

Suppose there were such a generalized measure corresponding to the
Kauffman bracket for some root of unity $q$ very close to 1.  If there
were, for some constant $C > 0$ the Kauffman bracket would satisfy
$|\langle K\rangle| < C$ for all framed knots $K$.  However, let $T$ be
the trefoil knot (with any framing).  Since $\langle T \rangle = (-A^5 -
A^{-3} + A^{-7})\langle \circ \rangle$ where $A = q^{1/4}$ is the
principal branch of the fourth root and $\circ$ denotes the unknot, for
$q$ sufficiently close to 1 we have $|\langle T\rangle | > |\langle
\circ \rangle|$.  On the other hand, for any two knots we have $\langle
K \# K'\rangle = \langle K \rangle \langle K' \rangle /\langle \circ
\rangle$, so by induction, the Kauffman bracket of a connected sum of
$n$ trefoil knots approaches infinity (in absolute value) as $n \to
\infty$, contradicting the supposed bound.

Alternatively, we can work with the Jones polynomial, an invariant of
oriented links arising from Chern-Simons theory with gauge group $\SU(2)
\times \U(1)$.  This does not depend on a framing, so {\it a priori} it
could arise from a generalized measure of the sort defined in this
section.  However, since the Jones polynomial of a knot $K$ is simply
the Kauffman bracket times $(-A^{-3})^{w(K)}$, where $w(K)$ is the
writhe, the above argument also shows that the Jones polynomial cannot
come from a generalized measure.  There is thus some mathematically
well-defined sense in which the Chern-Simons state is not
`normalizable.'  This does not yet rule it out as a physical state,
however, since it is possible that physical states of quantum gravity
can be more singular than generalized measures.

\section*{Acknowledgements}

I would like to Scott Axelrod, Matthias Blau, Paolo Cotta-Ramusino,
Louis Crane, James Dolan, and Maurizio Martellini for discussions
concerning the `dimensional ladder', and Abhay Ashtekar, Jerzy
Lewandowski, Jorge Pullin, Carlo Rovelli and Lee Smolin for discussions
on the loop representation.  I especially thank Steve Sawin for many
discussions, and for allowing me to publish his argument showing that
the Chern-Simons path integral does not come from a generalized measure.

\newcommand{\cqg}[1]{{\em Class.\ Quan.\ Grav.\ }{\bf #1}}
\newcommand{\grg}[1]{{\em Gen.\ Rel.\ Grav.\ }{\bf #1}}
\newcommand{\pr}[1]{{\em Phys.\ Rev.\ }{\bf #1}}
\newcommand{\prl}[1]{{\em Phys.\ Rev.\ Lett.\ }{\bf #1}}
\newcommand{\pl}[1]{{\em Phys.\ Lett.\ }{\bf #1}}
\newcommand{\np}[1]{{\em Nucl.\ Phys.\ }{\bf #1}}
\newcommand{\jmp}[1]{{\em J. Math.\ Phys.\ }{\bf #1}}
\newcommand{\jgp}[1]{{\em J. Geom.\ Phys.\ }{\bf #1}}
\newcommand{\cmp}[1]{{\em Commun.\ Math.\ Phys.\ }{\bf #1}}
\newcommand{\mpl}[1]{{\em Mod.\ Phys.\ Lett.\ }{\bf #1}}
\newcommand{\ijmp}[1]{{\em Int.\ J. Mod.\ Phys.\ }{\bf #1}}

\end{document}